\documentclass[12pt,a4paper]{amsart}
\pdfoutput=1
\usepackage{hyperref}
\usepackage{graphicx}
\usepackage{amsfonts}
\usepackage[T2A]{fontenc}
\usepackage[cp1251]{inputenc}
\usepackage{mathtext}
\usepackage{amsmath, amsfonts, amssymb}
\usepackage[english]{babel}
\usepackage[body={14.5cm, 20.5cm},left=4cm, top=4cm]{geometry}

\usepackage{epsf}
\usepackage{graphicx}

\def\be{\begin{eqnarray}}
\def\ee{\end{eqnarray}}
\def\bew{\begin{eqnarray*}}
\def\eew{\end{eqnarray*}}
\def\nn{\nonumber}

\def\p{\partial}

\def\Res{{\rm Res}\,}

\def\l[{\phantom.[}

\begin{document}

\bigskip

\centerline{\Large{\bf{Flat coordinates for Saito Frobenius manifolds and string theory}
}}

\bigskip

\bigskip

\centerline{{\bf A.~Belavin$^{a,c,e}$, D.~Gepner$^{d}$, Ya.~Kononov$^{a,b}$}}

\bigskip

{\footnotesize
\centerline{{\it
$^a$ L.\,D.~Landau Institute for Theoretical Physics, 142432 Chernogolovka, Russia}}

\centerline{{\it
$^b$ Higher School of Economics, Mathematics Department, Moscow, 117312, Russia
}}

\centerline{{\it
$^c$ Institute for Information Transmission Problems, Moscow, Russia
}}

\centerline{{\it
$^d$ Department of Particle Physics, Weizmann Institute, Rehovot, Israel
}}

\centerline{{\it
$^e$ Moscow Institute of Physics and Technology, Dolgoprudny, Russia
}}

\section*
{\bf{Abstract}}

It was shown in \cite{DVV} for $2d$ topological conformal field theory (TCFT) \cite{EY,W}
and more recently in \cite{BSZ}-\cite{BB2} for the noncritical string theory \cite{P}-\cite{BAlZ}
that several models of these two types can be exactly solved using their connection
with the Frobenius manifold (FM) structure introduced by Dubrovin \cite{Dub}.
More precisely, these models are connected with a special case of FMs, the so-called
Saito Frobenius manifolds (SFMs), originally called the flat structures  \cite{Saito}, 
which arise on the space of the versal deformations of the isolated singularities together 
with the flat coordinate systems after a suitable so-called primitive forms are chosen. 
In this paper, we explore the connection of the models of TCFT and noncritical string
theory with the SFM. The crucial point for obtaining an explicit expression for the
correlators is finding the {\it flat coordinates} of SFMs as functions of the parameters
of the deformed singularity.
We propose a new direct way to find the flat coordinates using the integral representation
for the solutions of the Gauss--Manin system connected with the corresponding SFM.
We illustrate our approach in the $A_n$ singularity case.
We also address the possible generalization of our approach for the models investigated
in \cite{Gep}, which are $SU(N)_k/(SU(N-1)_{k+1}\times U(1))$ Kazama--Suzuki theories \cite{KS}.
We prove a theorem that the potential connected with these models is an isolated singularity,
which is a condition for the FM structure to emerge on its deformation manifold.
This fact allows using the DVV approach to solve such Kazama--Suzuki models.

\bigskip

\bigskip

\section{\bf{Introduction}}

It is known that space--time supersymmetry in string theory
after \cite{Gep1,Banks} and before \cite{BelSpo} its compactification
is achieved because of the presence of the $N{=}2$ superconformal
symmetry on the worldsheet.
Moreover, chiral states and Ramond vacua of $N{=}2$ representations play
a crucial role in constructing massless-state operators.
It was shown in \cite{EY} that a twisted restriction to the chiral states
yields a topological conformal field theory (TCFT), which illustrates why
TCFT is important in string theory. The $N{=}2$ supersymmetric conformal
field theory (SCFT), and hence TCFT, was classified, somewhat generally,
in terms of the representation theory of affine Lie algebras in \cite{KS}.
The connection between the $N{=}2$ SCFT and singularity theory
\cite{Arnold, Blok} was investigated in \cite{Mart,VaWa,LVW}.
In \cite{Gep2}, a relation was established between the chiral rings of the
Kazama--Suzuki models $SU(N)_k/(SU(N-1)_{k+1}\times U(1))$ with
singularities, and the superpotential was shown to be a power-sum polynomial
expressed in terms of elementary symmetric polynomials.

Solving the mentioned theories explicitly has great interest. The important
series of minimal TCFT (or $SU(2)_k/U(1)$ theory) was solved in the seminal
paper by Dijkgraaf, Verlinde, and Verlinde (DVV) \cite{DVV}. These models
correspond to the so-called Gepner models of string compactification. These
models were solved exactly because they are connected with the first
nontrivial example of a Frobenius manifold (FM). (The concept of an FM
itself together with its axioms was introduced in the work of
Dubrovin \cite{Dub}.)

In addition to string theory in the space--time of the critical dimension
$d=26$ for bosonic and $d=10$ for fermionic string theories, there exists
another class of string theory models, the so-called noncritical strings
proposed by Polyakov \cite{P}. One class of such theories is the minimal
string theory or, in other words, minimal Liouville gravity (MLG) \cite{KPZ}.
Several MLG models were recently solved exactly \cite{BSZ,BDM,VB,BB1,BB2,BYR,BT,SPO}.
The exact solution of these models used their connection with Douglas
string equation \cite{Douglas} and with FMs \cite{BDM}.
It turned out that the generating function for the MLG correlators is the
logarithm of the Sato $\tau$-function, which is defined by a special
solution of the Douglas string equation. It is important that this
$\tau$-function can be expressed explicitly in terms of the structure
constants of the FM and the ``Douglas string action'' \cite{Ginsparg}.
The appropriate solution of the Douglas string equation satisfying the
selection rules has an adequate and simple expression in {\it flat
coordinates} \cite{VB, BB1}.

The above discussion shows that the problem of finding explicit expressions
for the flat coordinates of the corresponding FMs is
important for the exact solution of the above mentioned models. The theory of
FMs corresponding to the deformed singularities was
investigated in \cite{Saito, Noumi, Saito-Chan,Saito-Polvect}.

In this paper, we describe the connection between the exact solution of the
models of TCFT and $2d$ gravity connected with some isolated singularities
and the problem of computing the flat coordinates on FMs
associated with the same singularities.

\bigskip

In Sec.~2, we review the Dubrovin axioms and important properties of the
FM \cite {Dub}. We then describe a special class
of FMs connected with the versal deformations of isolated
singularities. We call such FMs {\it Saito Frobenius manifolds} (SFMs).
Namely, SFMs appear in the exactly solvable cases of TCFT and $2d$ gravity.
We also review the relation between FMs and the exactly solved models of
TCFT \cite {DVV} and $2d$ gravity \cite{BSZ}-\cite{BB1}. We also emphasise
that to obtain the correlators in both theories, we must know the
expressions for the flat coordinates as functions of the deformation
parameters of the corresponding isolated singularity.

In Sec.~3, we discuss the relation between the Jacobi ring of the deformed
isolated singularity and a Gauss--Manin system \cite{Saito,Noumi} connected
with the given Jacobi ring.

In Sec.~4, we show how to use the solutions of the Gauss--Manin system to
compute the flat coordinates. We illustrate our approach for directly
computing the flat coordinates in the case of the simple singularity $A_n$.
Our computation simplifies the previous computation by Noumi \cite{Noumi}.
The approach suggested here can be naturally generalised to other cases of SFM.

In Sec.~5, we discuss the possible generalization of our approach to the
case of the deformed $SU(N)_k/(SU(N-1)_{k+1}\times U(1))$ chiral
rings \cite{Gep} and prove a theorem on the isolatedness of the
corresponding singularity.

In the appendix, we show how to use the Gauss--Manin system and its
compatibility equations to prove the flatness of the metric defined as
$g_{\alpha\beta}=\Res\frac{e_\alpha e_\beta\zeta}{\prod\p_iF}$ in case of
the simple singularity. Here, $e_\alpha$ is an element of the Jacobi ring,
$F$ is the deformed singularity, and $\zeta$ is the so-called primitive
form \cite{Saito}.

\section{\bf{Background}}

\subsection{Frobenius manifolds (Dubrovin \cite{Dub})}

An FM $M$ is a Riemann manifold with coordinates $ t^0,t^2,\ldots,t^{\dim M-1}$,
a flat metric $g_{\alpha \beta}$, and a tensor $C_{\alpha \beta}^\gamma$ whose
components are the structure constants of an associative Frobenius algebra
identified with the tangent space of $M$,
\be
\frac{\p}{\p t^\alpha}\to e_\alpha,\\
e_\alpha e_\beta = C_{\alpha \beta}^\gamma e_\gamma.
\ee
The Riemann and Frobenius structures must be consistent as follows:
$g_{\alpha\beta}=(e_\alpha,e_\beta)$ is invariant, which means that
\be
(e_\alpha e_\beta,e_\gamma)=(e_\alpha,e_\beta e_\gamma)
\ee
and
\be
\nabla_\alpha C_{\beta\gamma}^\mu=\nabla_\beta C_{\alpha\gamma}^\mu,
\ee
where $\nabla_\alpha$ is the Levi-Civita connection for the flat metric
$g_{\alpha\beta}(t)$.
It then follows from associativity and Eq.~(4) that there exists a
one-parameter deformation of the Levi-Civita connection $\nabla_\alpha$ that
is also flat:
\be
\nabla_\alpha\quad\to\quad\tilde\nabla_\alpha=\nabla_\alpha+zC_{\alpha*}^*,
\ee
\be
[\tilde\nabla_\alpha,\tilde\nabla_\beta]=0.
\ee
It follows from (6) that there exist flat coordinates
$\theta^\mu(t^\alpha,z)$ for the deformed connection $\tilde\nabla_\alpha$
that can be expanded as a series in $z$:
\be
\theta^\mu(t^\alpha,z)=\sum_{k=0}^{\infty}\theta_k^\mu(t^\alpha)z^k.
\ee
The requirement that $\theta^\mu(t^\alpha,z)$ be deformed flat coordinates
can be written as
\be
\nabla_\alpha\nabla_\beta\theta_{k+1}^\mu=
C_{\alpha\beta}^\gamma\nabla_\gamma\theta_{k}^\mu.
\ee
As shown in \cite{BDM}, the coefficients $\theta_k^\mu$ of the expansion of
the flat coordinates can be used to construct the ``string action'' \cite{Ginsparg}.
We use this fact in Sec.~2.4.

\subsection{Saito Frobenius manifolds}

Let $f(x_1,...,x_n)$ be a quasihomogenous superpotential of an isolated
singularity, $e_\alpha$, $\alpha=0,\ldots,\dim M$, be the basis elements of
its Jacobi ring, and $F(x_i;t^{\alpha})$ be its versal deformation.
\be
F=f+\sum_{\alpha=0}^{\dim M-1}t^{\alpha}e_\alpha.
\ee
We use Latin indices like $i$ for the coordinates of the potential of the
singularity $f$ and Greek indices like $\alpha$ for its deformation
parameters.

As proposed by Saito \cite{Saito}, there then exists a form of the top
degree $\zeta$, the so-called primitive form, such that the metric defined
via $\zeta$ as
\be
g_{\alpha\beta}=\Res\frac{e_\alpha e_\beta\zeta}{\prod\p_iF}
\ee
is flat and, together with the structure constants
\be
C_{\alpha\beta\gamma}=\Res\frac{e_\alpha e_\beta e_\gamma\zeta}{\prod\p_iF},
\ee
provides an FM structure. In particular, the primitive forms for simple
singularities are given up to a constant factor by the standard volume form
$dx$. The local existence of a primitive form in the general case was proved
in \cite{M.Saito}. An explicit construction of weighted homogeneous
primitive forms $\zeta$ was given in \cite{Saito-Chan},\cite{Saito-Polvect}.

\subsection{TCFTs and the flat coordinates of SFMs}

Eguchi and Yang \cite{EY} showed that the TCFT can be obtained from the
$N{=}2$ SCFT by twisting the energy--momentum tensor
\bew
T_{top}(z)=T^{N=2}(z)+\frac{1}{2}\p J^{N=2}
\eew
and restricting the space of local fields of the $N{=}2$ SCFT model to its
subspace consisting of the chiral fields and their superpartners.

The symmetry of the TCFT is generated by the operators $L_n$, $Q_n$, $G_n$,
and $J_n$, whose commutation relations are
\bew
[L_m,L_n]=(m-n)L_{m+n},
\eew
\bew
[J_m,J_n]=dm\delta_{m+n,0},
\eew
\bew
[L_m,G_n]=(m-n)G_{m+n},
\eew
\bew
[J_m,G_n]=-G_{m+n},
\eew
\bew
[L_m,Q_n ]=-nQ_{m+n},
\eew
\bew
[J_m,Q_n]=Q_{m+n},
\eew
\bew
\{G_m,Q_n\}=L_{m+n}+nJ_{m+n}+\frac{1}{2}dm(m+1)\delta_{m+n,0},
\eew
\bew
[L_m,J_n]=-nJ_{m+n}-\frac{1}{2}dm(m+1)\delta_{m+n,0},
\eew
\bew
\{Q_m,Q_n\}=0.
\eew
Here, we use the braces in the case where both generators are odd and use
the square brackets otherwise. These relations include
\bew
L_n =\{Q,G_n\},\quad Q^2=0,\quad Q=Q_0,
\eew
which is exactly the definition of the TCFT \cite{DVV,EY,W}.

In the case of twisting $N{=}2$ minimal models connected with the coset
$SU(2)_k/U(1)$, the $\Phi_\alpha$ are subject to the constraints
\bew
\begin{cases}
Q\Phi_\alpha =0,\\
G_0\Phi_\alpha =0,\\
J_0\Phi_\alpha=q_\alpha\Phi_\alpha.
\end{cases}
\eew

In addition to $\Phi_\alpha$, the correlators in the TCFT also include their
superpartners $\tilde\Phi_\alpha$,
\bew
\tilde\Phi_\alpha=G_{-1}\bar G_{-1}\Phi_\alpha.
\eew
It follows from the above commutation relations that
\bew
Q\tilde\Phi_\alpha=\p\bar G_{-1}\Phi_\alpha\quad\Rightarrow\quad Q\int\tilde\Phi_\alpha=0.
\eew
All the information about the correlators of the theory is contained in the
perturbed correlators \cite{DVV}
\bew
F(s)_{\alpha_1,\alpha_2,\ldots,\alpha_r}=
\langle\Phi_{\alpha_1}(z_1)\cdots\Phi_{\alpha_r}(z_r)
e^{\sum s^\beta\int\tilde\Phi_\beta}\rangle,
\eew
which are independent of the coordinates $z_1,\ldots$ because
\bew
\p\Phi_\alpha=QG_{-1}\Phi_\alpha\quad\Rightarrow\quad
\p_i\langle\Phi_{\alpha_1}\cdots\Phi_{\alpha_m}\rangle=0.
\eew

It was shown in \cite{DVV} that all the information about
$F(s)_{\alpha_1,\alpha_2,\ldots,\alpha_r}$ is known if we know two of them:
\bew
C_{\alpha\beta\gamma}(s)=\langle\Phi_\alpha\Phi_\beta\Phi_\gamma
e^{\sum s^\delta\int\tilde\Phi_\delta}\rangle,
\eew
\bew
\eta_{\alpha\beta}=C_{\alpha\beta0}=
\langle\Phi_\alpha\Phi_\beta e^{\sum s^\delta\int\tilde\Phi_\delta}\rangle.
\eew
As proved in \cite{DVV}, $\eta_{\alpha\beta}$ and $C_{\alpha\beta\gamma}(s)$
satisfy the constraints that $\eta_{\alpha\beta}$ is independent of
$s^\rho$, $C_{\alpha\beta\gamma}(s)$ is a symmetric tensor, and
\bew
\p_\delta C_{\alpha\beta\gamma}=\p_\alpha C_{\delta\beta\gamma},
\eew
\bew
C_{\alpha\beta}^\rho C_{\rho\gamma}^\delta=
C_{\alpha\rho}^\delta C_{\beta\gamma}^\rho.
\eew
These relations show that we obtain an FM structure \cite{Dub}.

The $C_{\alpha\beta}^\rho (s)$ are structure constants of a ring $R$ that is
a deformation of the chiral ring of the minimal model. The coupling
constants $s^\alpha$ in the definition of the perturbed correlators are the
flat coordinates on the FM, and $\eta_{\alpha\beta}$ is the metric in the
flat coordinates.

Dijkgraaf, Verlinde, and Verlinde also proved \cite {DVV} that the perturbed
ring $R$ in the case of $SU(2)_k/U(1)$ minimal models coincides with the
ring of the versal deformation of the singularity $W_0=x^{k+2}$:
\be
\mathbb{C}[x]/\p_xW,
\ee
where the perturbed singularity is
\be
W=x^{k+2}+\sum_{\alpha=0}^kt_\alpha x^\alpha.
\ee
Therefore, the coupling constants $s^\alpha=s^\alpha(t_1,\ldots,t_k)$ are
functions of the perturbation parameters $t_1,t_2,\ldots\,$. They are just the
{\it flat coordinates} for the FM of the ring for the perturbed singularity
$\mathbb{C}[x]/\p_xW$. This implies that
\be
C_{\alpha\beta\gamma}(s)=\frac{\p t_\rho}{\p s^\alpha}
\frac{\p t_\mu}{\p s^\beta}\frac{\p t_\nu}{\p s^\gamma}
\Res_{x=\infty}\frac{x^{\rho+\mu+\nu}dx}{\p_x(x^{k+2}+\sum t_\delta x^\delta)}.
\ee
We recall that the primitive form in this case is just equal to $dx$. In
fact, this formula gives the exact solution for $SU(2)_k/U(1)$ minimal
models of TCFT.

Therefore, to obtain explicit expressions for the correlation functions of
the minimal models, we need only know the {\it flat coordinates} $s^\alpha$
as functions of the deformation parameters $t_1,t_2,\ldots\,$.

\subsection{MLG and flat coordinates on SFM}

The connection of MLG with the FM structure was found in \cite{BSZ,BDM} and
was used to construct the exact expressions for the correlation numbers of
the observables $O_{mn}$ \cite{BDM} of $(pq)$-MLG. Instead of the
correlators of the observables $O_{mn}$, it is convenient to study their
generating function
\be
Z^{MG}(\lambda)=\left\langle\exp\left(\sum_{m,n}
\lambda_{m,n}O_{m,n}\right)\right\rangle.
\ee
The result of the construction for $Z^{MG}(\lambda)$ in \cite{BDM} is then,
briefly, as follows.

Let $M$ be the FM connected with the versal deformation of the singularity
$x^{q+1}$,
\be
W(x;t_a)=x^{q+1}+\sum_{\alpha=0}^{q-1}t_ax^\alpha,
\ee
$g_{\alpha\beta}(t)$ be a metric on $M$, and $C_{\alpha\beta}^\gamma $ be
the structure constants of the perturbed ring defined as
\bew
g_{\alpha\beta}=\Res\frac{x^{\alpha+\beta}dx}{\p_xW(x;t)},
\eew
\bew
C_{\alpha \beta}^\gamma=g^{\gamma\delta}
\Res\frac{x^{\alpha+\beta+\delta}dx}{\p_xW(x;t)}.
\eew

We introduce the Douglas string equation \cite{Douglas}
\be
\frac{\p S(t,\tau_{mn})}{\p t^\alpha}=0,
\ee
in the formulation \cite{Ginsparg} in the form of constraints for the
critical points of the so-called $(p,q)$ {\it string action},
\be
S(t_\alpha,\tau_{mn})=\Res\left(W(x,t)^{p/q+1}+
\sum_{m,n}\tau_{m,n}W(x,t)^{|pm-qn|/q}\right)dx\nn,
\ee
where the residues are taken at $x=\infty$. We let $t_\alpha^*(\tau_{mn})$
denote the solutions of string equation~(17).

The string equation has a few solutions. The number of solutions depends on
$(p,q)$. For each solution $t^*(\tau_{mn})$ of the string equation, we can
define a function $Z(\tau_{mn})$ as \cite{BDM}
\be
Z(\tau_{mn})=\frac{1}{2}\int_0^{t^*(\tau_{mn})}C_\alpha^{\beta\gamma}
\frac{\p S}{\p s^\beta}\frac{\p S}{\p s^\gamma}dv^\gamma.
\ee
The function $Z (\tau_{mn})$ is the $\log\tau$-function of some integrable
hierarchy \cite{Dub},\cite{Krichever}.

Not every solution of the string equation is suitable for MLG, but there
does exist a suitable solution of the string equation and a choice of the
so-called resonance transformation from ``KdV'' frame variables $\tau_{mn}$
to the Liouville frame variables $\{\lambda_{m,n}\}$ such that
$Z(\tau(\lambda))$ coincides with the generating function of the $(p,q)$-MLG
$Z^{MG}(\lambda)$ \cite{BSZ, BDM,VB,BB1}.

It was found in \cite{VB} (also see \cite{BB1}) that the suitable solution
of the string equation has a very simple form in terms of the {\it flat
coordinates} of the FM, which is useful for the effective analysis of the
problem.

\section{\bf{Singularity, Gauss--Manin system, and Saito Frobenius manifold structure}}
We consider the case of a simple singularity defined by a quasihomogeneous
holomorphic function $f(x)\in\mathbb{C}[x_i]$. That the function $f$ is
quasihomogeneous means that there are weights $\nu_i$ such that
\be
f(\lambda^{\nu_i}x_i)=\lambda f(x_i).
\ee
The Milnor ring is defined by
\be
R_0=\mathbb{C}[x]/(\p_1f,\ldots,\p_nf).
\ee

We choose a quasihomogeneous basis $e_\mu$ with weights $[e_\mu]$ such that
$e_0=1$ for $R_0$ and consider a deformation
\be
F(x_i,t_\alpha)=f-t_0e_0-t_\alpha e_\alpha,\\
R=\mathbb{C}[x]/(\p_1F,\ldots,\p_nF).
\ee
Taking the weights $\epsilon_\mu$, of the parameters $t_\mu$, to be equal to
$1-[e_\mu]$, we ensure that $F$ is also quasihomogeneous,
\be
F(\lambda^{\nu_i}x_i,\lambda^{\epsilon_\mu}t_\mu)=\lambda F(x_i,t_\mu).
\ee

We define tensors $C$, $A$, and $B$ as
\be
e_\alpha e_\beta=C_{\alpha\beta}^\gamma e_\gamma+A_{\alpha\beta}^i\p_iF,\\
\sum_i\p_iA_{\alpha\beta}^i=-B_{\alpha\beta}^\gamma e_\gamma.
\ee
Because the singularity is simple (all weights $e_\mu<1$), we do not need to
add a term proportional to the derivatives in the second equation.

We define
\be
\Psi_0^{\mu}=\int_{C_\mu}\frac{dx}{F},
\ee
where the contours are defined to be eigenfunctions with the same monodromy
as $t^\mu$ and
\be
\Psi_\alpha^{\mu}=\p_{t_0}^{-1}\p_{t_\alpha}\Psi_0^{\mu},
\ee
which satisfies the so-called Gauss--Manin system of equations
\be
\p_{t_\alpha}\Psi_\beta^{\mu}=
C_{\alpha\beta}^\gamma\p_{t_0}\Psi_\gamma^{\mu}+
B_{\alpha\beta}^\gamma\Psi_\gamma^{\mu}.
\ee
We note that it follows from the definition of the tensors $C$, $A$, and $B$
that they are independent of $t_0$.

It is convenient to rewrite the Gauss--Manin system as a matrix differential
equation for the column $\Psi^{\mu}$ whose components are
$\Psi_\gamma^{\mu}$:
\be
{\mathcal D}_\alpha\Psi^{\mu}=(\p_\alpha-C_\alpha\p_0-B_\alpha)\Psi^{\mu}=0,
\ee
where $C_\alpha$ and $B_\alpha$ are matrices whose elements are equal to
$C_{\alpha\beta}^\gamma$ and $B_{\alpha\beta}^\gamma$.

If there are sufficiently many linearly independent solutions of the system
of equations, then we obtain the compatibility equations for the
Gauss--Manin system
\be
[{\mathcal D}_\alpha,{\mathcal D}_\beta]=0.
\ee
Because the matrices $C_\alpha$ and $B_\alpha$ are independent of $t_0$, we
obtain
\be
[C_\alpha,C_\beta]=0,
\ee
\be
\p_\alpha C_\beta-\p_\beta C_\alpha-[C_\alpha,B_\beta]+[C_\beta,B_\alpha]=0,
\ee
\be
\p_\alpha B_\beta-\p_\beta B_\alpha-[B_\alpha,B_\beta]=0.
\ee

Equation (33) for $B_{\alpha\beta}^\gamma$, together with the relation
\be
B_{\alpha\beta}^\gamma=B_{\beta\alpha}^\gamma,
\ee
shows that $B_{\alpha\beta}^\gamma$ can be regarded as Kristoffel
coefficients for the Levi-Civita connections of some flat metric. Taking
this fact into account, we conclude that Eqs.~(31)--(33) are just the
Dubrovin axioms for an FM structure.

Many checks support the {\it conjecture} that $B_{\alpha\beta}^\gamma$ is
identical to the Kristoffel coefficients for the Levi-Civita connections of
the metric
\be
g_{\alpha\beta}=\Res\frac{e_\alpha e_\beta}{\p_xF(x;t)}.
\ee
We prove the conjecture in the one-dimensional case in the appendix.

\section{\bf Gauss--Manin system and the flat coordinates}

\subsection{Action of the monodromy group on the solutions of the
Gauss--Manin system}

The integral representation of the solutions of the Gauss--Manin system can
be used to obtain explicit expressions for the flat coordinates $s^\mu$ in
terms of the deformation parameters $t_\alpha$.

In this section, we explain our approach in the case of the simple $A_n$
singularity. Noumi previously computed the flat coordinates in this
case \cite{Noumi}, but the approach suggested here simplifies this
computation and can be used in the general SFM cases. For this computation,
it is convenient to choose a special fundamental system of the solutions of
the Gauss--Manin system. Namely, we use the solutions that are
eigenfunctions under the monodromy group of transformations, whose action is
defined as $x_i\to\exp(i2\pi\nu_i)x_i$, $t_0\to t_0$, and
$t_\mu\to\exp(i2\pi\epsilon_\mu)t_\mu$. The Gauss--Manin system is invariant
under such a transformation.

Therefore, we can choose our fundamental system of the solution $\Psi^{\mu}$
such that
\be
\Psi_0^{\mu}(t_0,\exp(2\pi i\epsilon_\beta)t_\beta)=
\exp(2\pi i\epsilon_\mu)\Psi_0^{\mu}(t_0,t_\beta)
\ee
and
\be
\p_0\Psi_\alpha^{\mu}(t_0,t_\sigma)=\p_\alpha\Psi_0^{\mu}(t_0,t_\sigma).
\ee
It follows from the first of these equations that
\be
\Psi_0^{\mu}(t_0,t_\beta)=
t_0^{\epsilon_\mu}G\left(\frac{t_\beta}{t_0^{\epsilon_\beta}}\right).
\ee
If the combinations $t_\beta/t_0^{\epsilon_\beta}$ are small, then we have
an expansion
\be
\Psi_0^\mu(t_0,t_\sigma)=
\sum_{N=0}\frac{U^\mu(t_\beta,N)}{t_0^{\epsilon+N}},
\ee
where
\be
U^{\mu}(t_\beta,N)=\sum_{m_\alpha:\,\sum\epsilon_\alpha m_\alpha=\epsilon_\mu+N}
C^{\mu}(m_\beta)\left(\prod \frac{t_\alpha^{m_\alpha}}{m_\alpha!}\right).
\ee
and 
$$\epsilon=1-\sum \nu_i$$

Similarly, we obtain an expansion for $\Psi_\alpha^\mu(t_0,t_\sigma)$,
\be
\Psi_\alpha^\mu(t_0,t_\sigma)=\sum_{N=0}\frac{U_\alpha^\mu(t_\beta,N)}
{t_0^{\epsilon+N}},
\ee
and it follows from Eq.~(37) that
\be
U_\alpha^{\mu}(t_\sigma,N=0):=
U_\alpha^{\mu}(t_\sigma)=\p_\alpha U_0^{\mu}(t_\sigma,N=0).
\ee

\subsection{Solutions of the Gauss--Manin system and flat coordinates}
The structure constants $C_{\alpha\beta}^\gamma$ and $B_{\alpha\beta}^\gamma$
are independent of $t_0$. It follows from this fact and Eqs.~(28) and (41)
that
\be
(\p_\alpha-B_\alpha)U_\beta^\mu(t_\sigma)=0.
\ee
This means that
\be
U_\alpha^\mu(t_\sigma)=\frac{\p s^\mu}{\p t^\alpha},
\ee
where the functions $s^\alpha(t_\beta)$ are the flat coordinates.

Comparing Eqs.~(44) and (42), we conclude that the flat coordinates coincide
with the leading terms for $\Psi_0^\mu(t_0,t_\sigma)$ in expression (39),
i.e.,
\be
s^\mu(t_\sigma)=U_0^{\mu}(t_\sigma).
\ee
With the change of coordinates $t^\alpha \to s^\alpha$, a direct computation
shows that
\be
B_{\alpha\beta}^\gamma\to\tilde B_{\alpha\beta}^\gamma=0,
\ee
\be
C_{\alpha\beta}^\gamma\to\tilde C_{\alpha\beta}^\gamma.
\ee
In the new coordinates, we have
\be
\p_{\alpha}\tilde C_{\beta\gamma\delta}=
\p_\beta\tilde C_{\alpha\gamma\delta}.
\ee
Taking into account that $\tilde C_{\beta\gamma\delta}$ are independent of
$s^0$, we obtain
\be
\p_\alpha\tilde C_{0\beta\gamma}=\p_0\tilde C_{\alpha\beta\gamma}=0.
\ee

Taking into account that $\tilde C_{0\alpha\beta}=\tilde g_{\alpha\beta}$,
we find that the metric $g_{\alpha\beta}$ defined in (35) in the new
coordinates is constant.

We thus confirm our conjecture that $B_{\alpha\beta}^\gamma$ are Kristoffel
symbols for the Levi-Civita connection of $g_{\alpha\beta}$.

\subsection{The explicit calculation of flat coordinates: Example of the
$A_n$ singularity}
Here, we show how to use integral representation (26) for
$\Psi_0^\mu(t_0,t_\sigma)$ to obtain an explicit expression for the flat
coordinates. We consider the example of the versal deformation of the
one-dimensional $A_p$ singularity
\be
F=x^{p+1}-t_0+\sum_{1\leq\alpha\leq p-1}t_\alpha x^\alpha.
\ee

We choose the initial contours $C_n$ in the definition of $\Psi_0^{(n)}$ as
shown in Fig.~1:
\be
\Psi_0^{(n)}=\int_{C_n}\frac{dx}{F}.
\ee
In this case, the integral is taken around the pole in
\be
x^{(n)}=(t_0)^{1/(p+1)}\exp(2\pi in/(p+1))
\ee
if $t_\alpha$ are small for $\alpha>0$.

\begin{figure}[!htb]
\vspace{2mm}
\begin{center}
\includegraphics[height=2.20in, width=2.20in]{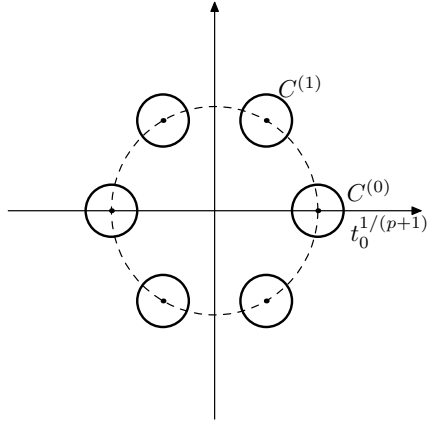}
\end{center}
\caption{Initial contours}
\label{fig}
\end{figure}

By definition, $\Psi_0^{(n)}(t_0,t_\beta)$ satisfy the relation
\be
\Psi_0^{(n)}(t_0,\omega^{(p+1-\beta)}t_\beta)=
\bar\omega\Psi_0^{(n+1)}(t_0,t_\beta),
\ee
where $\omega=\exp(2\pi i/(p+1))$. For our aim, instead of
$\Psi_0^{(n)}(t_0,t_\beta)$, it is convenient to choose their linear
combinations
\be
\Psi_0^{\mu}(t_0,t_\beta)=
\sum_n\bar\omega^{\mu n}\Psi_0^{(n)}(t_0,t_\beta),
\ee
which are eigenfunctions of the monodromic transformation
\be
\Psi_0^{\mu}(t_0,\omega^{(p+1-\beta)}t_\beta)=
\omega^{\mu} \Psi_0^{\mu}(t_0,t_\beta).
\ee
As explained above, with the combinations $t_\beta/t_0^{\epsilon_\beta}$,
where $\epsilon_\beta=(p+1-\beta)/(p+1)$ are small, we have the expansion
\be
\Psi_0^\mu(t_0,t_i)=
\sum_{N\geq0}\frac{U^\mu(t_i, N)}{t_0^{\epsilon+N}},
\ee
where
\be
U^{\mu}(t_\beta,N)=
\sum_{m_\alpha:\,\sum\epsilon_\alpha m_\alpha=\epsilon_\mu+N}
C^{\mu}(m_\alpha)\left(\prod\frac{t_\beta^{m_\beta}}{m_\beta!}\right).
\ee

To compute $\Psi_0^\mu(t_0,t_\beta)$, we use integral representation~(51)
for $\Psi_0^{(n)}(t_0, t_\beta)$. Indeed, we do not need to compute the
integrals along all contours $C^{(n)}$ and take their linear combination
defined in (54). The weighted summation in (54) plays only a role of the
projector of each of the integrals along the contours $C^{(n)}$ to the part
that corresponds to the eigenfunction with a definite monodromic behaviour.
We therefore concentrate on computing one of the integrals, namely, along
the contour $C^{(0)}$ and in it retain only those terms of its series that
have the needed monodromic behaviour.

We deform the contour $C^{(0)}\to\tilde C^{(0)}$ as shown in Fig.~2. This
deformation of the contour allows performing an integral representation for
the integrand $1/F$ taking into account that
\be
f(x)=x^{p+1}-t_0
\ee
has a negative real part on the contour.

\begin{figure}[!htb]
\vspace{2mm}
\begin{center}
\includegraphics[height=2.20in, width=2.20in]{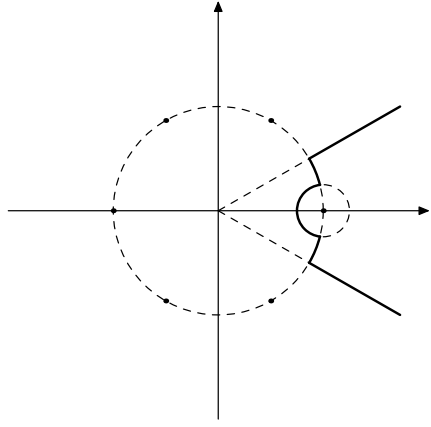}
\end{center}
\caption{Deformation of the contour $C^{(0)}\to \tilde C^{(0)}$}
\label{fig}
\end{figure}

It follows that we can use the integral representation
\be
F^{-1}=\int_0^\infty dy\,e^{Fy}.
\ee
We assume that the variables $t_{\geq1}$ are small, as explained above.
Therefore, we can perform the Taylor expansion in $t_\alpha$ with
$\alpha\geq1$:
\begin{multline}
\Psi_0^{(0)}=\int_{\tilde C^{(0)}}dx\int _0^\infty dy\,e^{y(x^{p+1}-t_0+
\sum t_\alpha x^\alpha)}=\\
=\int_{\tilde C^{(0)}}dx\int_0^\infty dy\,e^{yx^{p+1}}e^{-t_0y}
\sum_{m_\alpha}\left(\prod\frac{t_\alpha^{m_\alpha}}{m_\alpha!}\right)
x^{\sum\beta m_\beta}y^{\sum m_\gamma}.
\end{multline}
Changing the integration order and changing the variables
\be
x\to y^{-1/(p+1)}x,
\ee
we obtain
\begin{multline}
\Psi_0^{(0)}=\int dy\,y^{\sum m_\alpha-\frac{\sum\alpha m_\alpha+1}
{p+1}}e^{-t_0y}\left(\prod\frac{t_\beta^{m_\beta}}{m_\beta!}\right)
\int_{\tilde C^{(0)}}dx\,e^{x^{p+1}}x^{\sum\beta m_\beta}=\\
=\sum_{m_\alpha}\frac{\Gamma\left(\sum m_\alpha-
\frac{\sum\alpha m_\alpha+1}{p+1}+1\right)}
{t_0^{\sum m_\alpha-\frac{\sum\alpha m_\alpha+1}{p+1}+1}}
\left(\prod\frac{t_\beta^{m_\beta}}{m_\beta!}\right)
\int_{\tilde C^{(0)}}dx\,e^{x^{p+1}} x^{\sum\beta m_\beta}.
\end{multline}

In the $x$ integral, we can now deform the contour to two rays passing
through the origin with the angles $\pm\pi/(p+1)$, and a direct calculation
then shows that
\be
\int_{\tilde C^{(0)}}dx\,e^{x^{p+1}}x^r=
\frac{2 i}{p+1} \sin \pi \left (\frac{ r+1}{p+1}\right) \Gamma\left(\frac{r+1}{p+1}\right)
\ee
We introduce $\mu$ and $N$ depending on $m_i$,
\be
\sum m_\alpha \epsilon_\alpha= \epsilon_\mu +N,
\ee
where $0\leq\mu\leq (p-1),  0\leq N$.

Then
\begin{multline}
\Psi_0^{(0)}=\frac{2 i}{p+1}\sum_{\atop  0\leq\mu\leq (p-1)}\sum_{N\geq0}
\frac{\Gamma(\epsilon_\mu+\epsilon+N)}{t_0^{\epsilon_\mu+\epsilon+N}}
\sum_{m_\alpha \atop \sum m_\alpha \epsilon_\alpha= \epsilon_\mu +N}
\left(\prod\frac{t_\alpha^{m_\alpha}}{m_\alpha!}\right)
\sin \left (\pi  \frac{\sum\alpha m_\alpha+1}{p+1}\right) \cdot \\ \cdot
\Gamma\left ( \frac{\sum\alpha m_\alpha+1}{p+1}\right) 
\end{multline}

From this formula, as explained above, we now extract the expression for
$\Psi_0^{\mu}$:
\begin{multline}
\Psi_0^{\mu}=\frac{2 i}{p+1}
\sum_{N\geq0}
\frac{\Gamma(\epsilon_\mu+\epsilon+N)}{t_0^{\epsilon_\mu+\epsilon+N}} 
\sum_{m_\alpha \atop \sum m_\alpha \epsilon_\alpha= \epsilon_\mu} 
\left(\prod\frac{t_\alpha^{m_\alpha}}{m_\alpha!}\right) 
\sin \left (\pi \frac{\sum\alpha m_\alpha+1}{p+1}\right) 
\Gamma\left ( \frac{\sum\alpha m_\alpha+1}{p+1}\right) 
\end{multline}

The leading term in $\Psi_0^{\mu}$ with $N=0$ corresponds up to a normalization,
as was explained above, to the flat coordinates in terms of the deformation parameters.
Also we will take  into account that from eq. (64) for $N=0$   follows that
\be
\sum\alpha m_\alpha=\mu+k(p+1)
\ee
where $ k\geq0$.  

Using both these facts we obtain the explicit expression for the flat coordinates
\be
s^\mu=\sum_{m_\alpha \atop \sum m_\alpha \epsilon_\alpha= \epsilon_\mu }(-1)^{k}
\frac{\Gamma\left(\frac{\mu+1}{p+1}+k\right)}{\Gamma\left(\frac{\mu+1}{p+1}\right)}
\left(\prod\frac{t_\alpha^{m_\alpha}}{m_\alpha!}\right).
\ee

Integrating in (62), we use the well-known expression for the gamma function.

In general, we could use Stokes's theorem to reduce integrals of the form
\be
\int e^{f/z}\omega
\ee
defined on the cohomology space
\be
\omega\in\Omega^n[[z]]/(zd+df\wedge)\Omega^{n-1},
\ee
called the Brieskorn lattice. In other words,
\be
\int e^{f/z}\omega=\int e^{f/z}\omega'\quad\text{if}\quad
\omega-\omega'=z\,d\eta+df\wedge\eta\quad\text{for some}\quad
\eta\in\Omega^{n-1}.
\ee

Thus  in this section we have shown that the flat coordinates  in  the case of $ A_n $ singularity are given  by the leading term of the  asymptotic of the solution  of the Gauss-Manin system (28)   $\Psi_0^\mu(t_0,t_\sigma)$  which is defined by the  integral representation (26).

We expect that a similar approach can be used for coputation of the flat coordinates in the general case 
of the isolated singularity when not all weights  $ e_{\mu}<1 $. However the integral representation for the solution  of the Gauss-Manin system (28) in this case has to be modified as follows 

\be
\Psi_0^{\mu}=\int_{C_\mu}\frac{\zeta}{F},
\ee
where  the Saito primitive form $\zeta$ \cite{Saito} is inserted instead of the standart volume form $dx$.

\section{\bf Gepner singularities}

We consider some Kazama--Suzuki $N{=}2$ superconformal field theories that
are cosets of the type
\be
G=SU(N)_k\times SO(2N-2)_1/SU(N-1)_{k+1}\times U(1).
\ee
The primary fields of the model are labeled by representations of $SU(N)_k$,
denoted by $\Lambda$; a representation of $SO(2N-2)_1$, denoted by $r$; and
a representation of $SU(N-1)_{k+1}\times U(1)$, denoted by $(\lambda,q)$. We
write this field as
\be
H^{\Lambda,\rho}_{\lambda,q}={SU(N)^\Lambda_k\times SO(2N-2)_1^r/
SU(N-1)_{k+1}^\lambda\times U(1)_q}.
\ee
This theory has a $N{=}2$ superconformal symmetry, where the super stress
tensor is given by
\be
G_+(z)=\sum_{\alpha\in\Delta_{+,R}-\Delta_{+,U}}J^\alpha(z)\rho^{-\alpha}(z),
\ee
where $\Delta_{+,R}$ are the positive roots of $R=SU(N)$ and the same for
$U=SU(N-1)$.

The chiral fields are the primary fields $C$, which satisfy
\be
G_+(z)C(\zeta)={\rm regular},
\ee
\be
G_-(z)C(\zeta)=O(1/(z-\zeta)).
\ee
These fields satisfy $\Delta=q/2$, where $\Delta$ is the dimension and $q$
is the $U(1)$ charge. For the considered model $G$ given by Eq.~(73), it can
be shown that the chiral fields are in one-to-one correspondence with the
representations of $SU(N)_k$, denoted by $\Lambda$. The chiral fields can be
written as \cite{Gep2}
\be
C^\Lambda(z)=H^{\Lambda,0}_{\Lambda,q}(z),
\ee
where $\Lambda$ is any representation of $SU(N)_k$ and the $U(1)$ charge is
given by
\be
q=2(\rho-\hat\rho)\Lambda/(k+N),
\ee
where $\rho$ and $\hat\rho$ are half the sum of positive roots for $R$ and
$U$.

We now consider the OPE algebra of the chiral fields. The product of the two
fields $C^\Lambda$ and $C^\mu$ is given just by the fusion product of the
representations $\Lambda$ and $\mu$ in $SU(N)_k$ such that the $U(1)$ charge
is preserved \cite{Gep2}. It is convenient to express the representation as a Young
tableau or a sequence of integers $0<a_1\leq a_2\leq\dots\leq a_k\leq N-1$,
where the $a_i$ is the height of the $i$th column of the Young tableau,
$\Lambda=\sum_{i=1}^k\Lambda^{(a_i)}$, where $\Lambda^{(r)}$ is $r$th
fundamental weight. We have the fully antisymmetric representations
$\bar c_r=[0,0,\ldots,0,r]$ and the fully symmetric representations
$c_i=[1,1,\ldots,1]$.

Therefore, we need to study the fusion rings of $SU(N)_k$. These were
studied in \cite{Gep}. The generators of the ring are the fully antisymmetric
representations $\bar c_r$, whose $U(1)$ charges are $r/(k+N)$. The fusion
ring is described by a Pieri-like formula. When we impose the $U(1)$ charge
conservation, we obtain just the Pieri formula for the multiplication of
chiral fields,
\be
\bar c_r\times[a_1,a_2,\ldots,a_k]=
\sum_{a_i\leq b_i\leq a_{i+1}\atop r+\sum a_i=\sum b_i}
[b_1,b_2,\ldots,b_k],
\ee
where $\Lambda=[a_1,a_2,\ldots,a_k]$ denotes the field $C^\Lambda$. As a
side remark, we note that this is just the formula for multiplying Schubert
cells in the cohomology of the Grassmanian manifold
\be
G_k(C^{N+k-1})=U(N+k-1)/U(N-1)\times U(k).
\ee
The relations in the chiral ring can be shown to be given by the fully
symmetric representations $c_r$, $r=k+1,k+2,\ldots,k+N-1$. Namely, these
fields are equal to zero in the chiral ring. As shown in \cite{Gep}, these
relations can be derived from the potential
\be
V=q_1^{k+N}+q_2^{k+N}+\dots+q_{N-1}^{k+N}
\ee
expressed in terms of the fully symmetric functions,
\be
x_r=\bar c_r=q_1q_2\ldots q_r+{\rm symmetrise}.
\ee
We have thus proved that the chiral ring of the Kazama--Suzuki model $G$ is
isomorphic to the Jacobi ring with the superpotential $V$,
\be
{\mathcal R}=P[x_1,x_2,\ldots,x_n]/(\p_iV).
\ee

We now discuss fusion rings that are describable by a potential. For
example, $SU(N)_k$ fusion rings are known \cite{LVW} to be given by the potential
\be
V=q_1^{N+k}+q_2^{N+k}+\dots+q_N^{N+k}
\ee
with the constraint $q_1q_2\ldots q_N=1$ in terms of the symmetric functions
$x_i$ in the variables $q_1,q_2,\ldots,q_N$, which are the fully
antisymmetric Young tableaux $\bar c_i$. The points of the vanishing of the
derivatives of the potential, $\p_iV=0$, are called the fusion variety and
are given by
\be
x_i=S_{a,i}/S_{0,i}
\ee
for any rational conformal field theory (RCFT), where $S$ is the modular
matrix and $a$ is any of the primary fields labelling the points \cite{Gep}. This
is a collection of distinct points, or the fusion ring is a semisimple ring.

We can regard this theory as a massive $N{=}2$ conformal field theory,
namely, a nonquasihomogenous potential. This theory flows to a $N{=}2$ SCFT,
which is obtained by imposing the $U(1)$ charge conservation or
quasihomogeneity. We also assume that if the fusion coefficient
$f_{ijk}\neq0$, then $\nu_i+\nu_j\geq\nu_k$, where $\nu_i$ is the weight (or
$U(1)$ charge) and and also assume that the equality holds only for the $N{=}2$ SCFT. For
example, the $SU(N)_k$ fusion ring flows to the Kazama--Suzuki model $G$
given by Eq.~(73).

The flow to the conformal field theory can now be described by a parameter
$t$ such that $t=1$ is the fusion ring and $t=0$ is the superconformal
chiral ring and we transform each field as
\be
x_i\to t^{\nu_i}x_i.
\ee
Hence, at $t=0$, we obtain $x_i=0$ as the only solution. We have thus proved
that the singularity is isolated, which we express in the following theorem.

\medskip

{\bf Theorem.}
{\it Consider a chiral ring that is a limit of a fusion ring, namely, the
fusion coefficient $f_{ij}^k$ in the fusion ring is nonzero only if the
$U(1)$ charges satisfy $\nu_i+\nu_j\geq\nu_k$, where the equality holds only
for the chiral ring fields. Then such a ring is an isolated singularity.}

\medskip

In particular, we proved that the chiral ring $\mathcal R$ is an isolated
singularity, as was also shown in \cite{Gep,Witten}, and recently rigorously proved
by Saito \cite{Saito1}. In fact, this theorem also holds when the fusion ring is not
described by a potential, showing that the only zeros of the relations of
the chiral ring are the point zero, as long as it is such a limit of a
fusion ring of an RCFT. In other words, we proved that such chiral rings are
always local. The significance of this result is that it implies that such
rings satisfy the FM axioms.

\section {\bf{ Conclusion.}}

One of the most beautiful aspects of compactified string theory is the
interplay between algebraic constructions of string theory based on solvable
conformal field theories as opposed to the geometric sigma model approach.
A connection was established in \cite{Gep3} between the so-called Gepner
models of string theory and the Calabi--Yau compactifications. Many examples
of manifolds were found, thus giving an exact solution of the previously
abstruse Calabi--Yau compactification. Many more examples were later found
in \cite{Mart,VaWa,Gep2} based on a Landau--Ginzburg description of the
solvable Gepner string theories.

The moduli of the Calabi--Yau manifolds are of two kinds: those that change
the complex structure and those that change the radii. Both of these are
counted geometrically by the Hodge numbers $h^{2,1}$ and $h^{1,1}$. In the
conformal field theory, they appear as superpartners of the chiral and
antichiral fields with dimension $1/2$. Calculating physical quantities away
from the solvable point therefore involves perturbation theory calculations
with these fields. This is exactly the sort of calculation performed in
\cite{DVV}, except that the authors did not limit themselves to marginal
fields, i.e., the dimension-one fields. It is hence also of interest to
perform this sort of analysis in string compactification. We can speculate
that the flat coordinates correspond to the periods of the manifold. This is
hard to analyze because we do not know the metric in general, but toric
examples support this conjecture.

The Calabi--Yau manifolds corresponding to the $SU(N)_k/(SU(N-1)_{k+1}\times
U(1))$ Kazama--Suzuki models of TCFT are of particular interest. The
corresponding manifold was described in \cite{Gep2} by equating the
superpotential to $0$ as an equation in the weighted projective space.
Hence, it is also of interest in critical string theory to calculate the
deformations of these Kazama--Suzuki TCFTs.

Using the approach described in the preceding sections to solve the
$SU(N)_k/(SU(N-1)_{k+1}\times U(1))$ models is possible because they
correspond to isolated singularities. This follows because the property of
the isolation of the singularity, as follows from the results of Saito and
others \cite {Saito, M.Saito, Saito-Chan, Saito-Polvect}, is exactly the
condition for the emergence of the FM structure on the space of its
deformations. The FM structure, as we have emphasised in this paper, has a
crucial importance for solving such TCFTs.

Another interesting open problem is to investigate the models of $2d$
minimal W-gravities connected with $SU(N)_k/(SU(N-1)_{k+1}\times U(1))$
singularities as was done in \cite{BSZ}-\cite {BB1} in the $SU(2)_k/U(1)$
case.

\section{\bf{Acknowledgments}}
We are grateful to V.~Belavin, M.~Berstein, G.~Giorgadze, S.~Gusein-Zade,
V.~Gorbunov, K.~Hori, A.~Ionov, S.~Lando, Yu.~Manin, T.~Milanov, and
K.~Saito for the useful conversations. We thank Ida Deichaite and Bill Everett for the
remarks on the manuscript.

A.~B.~is grateful to K.~Narain and F.~Qevedo for the warm hospitality
at the ICTP, Trieste, and for the interesting discussions. Ya.~K.~ is
thankful to Professor K.~Saito for the hospitality at IPMU and to the Simons
Center. The research of A.~B.~ was performed at the IITP RAS and
supported by the RSF (Project No.~14-50-00150).

\section{Appendix: A note on the conjecture}

While investigating the Gauss--Manin connection, our natural assumption was
that $B_{\alpha\beta}^\gamma=\Gamma_{\alpha\beta}^\gamma$, where the latter
denotes the Kristoffel symbols for the Levi-Civita connection associated
with the metric $g_{\alpha \beta}$. This was based on the observation that
one of the compatibility equations was very similar to the flatness of the
connection. We need to verify the compatibility of $B_{\alpha\beta}^\gamma$
with the metric $g_{\alpha \beta}$. This property can be written as
\be
\p_\alpha g_{\beta\gamma}=
(\nabla_\alpha e_\beta,e_\gamma)+(e_\beta,\nabla_\alpha e_\gamma).
\ee
We consider the difference between the left- and right-hand sides:
\begin{multline*}
\p_\alpha g_{\beta\gamma}-(\nabla_\alpha e_\beta,e_\gamma)-
(e_\beta,\nabla_\alpha e_\gamma)=\Res\frac{e_\beta e_\gamma}{\prod_j\p_jF}
\frac{\p_ie_\alpha}{\p_iF}-\Res\frac{B_{\alpha\beta}e_\gamma}{\prod_j\p_jF}-
\Res\frac{B_{\alpha\gamma}e_\beta}{\prod_j\p_j F}=\\
=\Res e_\alpha e_\beta e_\gamma\sum_i\left(\frac{\p_i
\left(\frac{1}{\prod\p_jF}\right)}{\p_iF}-\frac{\p_i
\left(\frac{1}{\p_iF}\right)}{\prod\p_jF}\right).
\end{multline*}
In the one-dimensional case, the expression in the parentheses is identical
to zero.

\end{document}